\documentclass[12pt]{article}
\input epsf
\usepackage{alltt}
\renewcommand{\baselinestretch}{1.21}
\newcommand{\grant}[1]{\footnote{To Grant: #1}}
\newcommand{\kalvis}[1]{\marginpar{To Kalvis: #1}}
\newcommand{\note}[1]{\marginpar{NB: #1}}
\newcommand{\tN}{{\bf N}}
\newcommand{\tF}{{\bf F}}
\newcommand{\tB}{{\bf B}}
\newcommand{\tW}{{\bf W}}
\newcommand{\np}[1]{{\rm\bf n}_+\{ \mbox{#1} \}}
\newcommand{\nn}[1]{{\rm\bf n}_-\{ \mbox{#1} \}}
\newcommand{\npu}[1]{{\rm\bf n}_+^u \{ \mbox{#1} \}}
\newcommand{\nnu}[1]{{\rm\bf n}_-^u \{ \mbox{#1} \}}
\newcommand{\ou}{Ornstein--Uhlenbeck }
\newcommand{\npa}[1]{{\rm\bf n}_+^{(\alpha)} \{ \mbox{#1} \}}
\newcommand{\nna}[1]{{\rm\bf n}_-^{(\alpha)} \{ \mbox{#1} \}}
\newcommand{\cosech}{\, {\rm cosech} \,}
\newcommand{\xt}{{{\bf X}_t}}
\newcommand{\xtz}{{{\bf X}_{t_0}}}
\newcommand{\xs}{{{\bf X}_s}}
\newcommand{\xz}{{{\bf X}_0}}
\newcommand{\xx}{{{\bf X}}}
\newcommand{\yy}{{{\bf Y}}}
\newcommand{\ww}{{{\bf W}}}
\newcommand{\st}{{{\bf Q}_t^{(i)}}}
\newcommand{\stb}{{{\bf Q}_{\tb}^{(i)}}}
\newcommand{\xu}{\underline{\bf X}}
\newcommand{\xo}{\overline{\bf X}}
\newcommand{\bt}{{\bf W}_t}
\newcommand{\wt}{{\bf W}_t}
\newcommand{\ws}{{\bf W}_s}
\newcommand{\bs}{{\bf W}_s}
\newcommand{\bz}{{\bf W}_z}
\newcommand{\wz}{{\bf W}_z}
\newcommand{\yt}{{\bf Y}_t}
\newcommand{\ys}{{\bf Y}_s}
\newcommand{\za}{{\bf Z}_a}
\newcommand{\nt}{{\bf N}_t}
\newcommand{\ba}{{\bf a}}
\newcommand{\ha}{\half}
\newcommand{\ee}{{\rm e}}
\renewcommand{\d}{{\rm d}}
\newcommand{\gauss}[1]{\frac{{\rm e}^{-\frac{#1^2}{2 t}}}{(2 \pi t)^\half}}
\newcommand{\E}[1]{{\bf E}\left[ #1 \right]}
\newcommand{\ind}[1]{{\cal I}{\left\{{#1}\right\}}}
\newcommand{\pp}[2]{{\cal P}^{(#1)}{\left( #2 \right)}}
\newcommand{\pr}[1]{{\cal P}[#1]}
\newcommand{\tset}{{\cal I}}
\newcommand{\e}{{\rm e}}
\newcommand{\half}{{\mbox{\(\frac{1}{2}\)}}}
\newcommand{\qua}{{\mbox{\(\frac{1}{4}\)}}}
\newcommand{\pxt}{p_{\bf X}}
\newcommand{\eqref}[1]{(\ref{#1})}
\newcommand{\F}{ F}
\renewcommand{\L}{ L}
\newcommand{\ito}{It\^o}
\newcommand{\borel}{{\cal B}}
\newcommand{\tx}{\frac{\partial T}{\partial x}}
\newcommand{\txx}{\frac{\partial^{2} T}{\partial x^{2}}}
\newcommand{\tb}{{\bf t}_b}
\newcommand{\tz}{{\bf t}_0}
\newcommand{\tH}{{\bf t}_h}
\newcommand{\sss}{{\bf s}}
\newcommand{\ssb}{{\bf s}_b}
\newcommand{\lt}{{\bf L}_t}
\newcommand{\lx}{{\bf L}_x}
\newcommand{\la}{{\bf L}_a}
\newcommand{\laq}{[{\bf L}]_a}
\newcommand{\ltx}{{\bf L}_t(x)}
\newcommand{\lta}{{\bf L}_t(a)}
\newcommand{\ltba}{{\bf L}_{{\bf t}_b}(a)}
\newcommand{\gl}{{\bf g}_l}
\newcommand{\mle}{\mu|\log\epsilon|}
\newcommand\mean[1]{{\big<#1\big>}}
\newcommand{\rxt}{R(x)}
\newcommand{\eps}{\epsilon}
\newcommand{\var}{v}
\newcommand{\limvar}{w}
\newcommand{\gd}{q}
\newcommand{\iint}{{\bf I}_t}

\begin{document}

\begin{center}\LARGE\bf
Stochastic calculus:\\
application to dynamic bifurcations and threshold crossings
\end{center}
\begin{center}\large
        Kalvis M. Jansons\(^{1}\) and G.D. Lythe\(^{2,3}\)
\end{center}

\begin{center}
{\em To appear in Journal of Statistical Physics}
\end{center}
\begin{center}
\(^{1}\)Department of Mathematics, University College London, Gower 
Street, London WC1E 6BT, England\\
\(^{2}\)Optique nonlin\'eaire th\'eorique, Universit\'e Libre de Bruxelles
CP 231, Bruxelles 1050, Belgium.\\
\(^{3}\){\em current address:
Center for nonlinear studies, B258,\\
 Los Alamos National Laboratory, NM87544, USA.}
\begin{alltt}
\centerline{http://cnls-www.lanl.gov/~grant/}
\end{alltt}
\end{center}


\begin{abstract}
\noindent
For the dynamic pitchfork bifurcation in the presence of white noise,
the statistics of the last time at zero are calculated as a function
of the noise level, \(\eps\), and the rate of change of the parameter,
\(\mu\).
The threshold crossing problem used, for example, to model
the firing of a single cortical neuron is considered, concentrating on
quantities that may be experimentally measurable but have so far
received little attention.  Expressions for the statistics
of pre-threshold excursions, occupation density and last crossing time
of zero are compared with results from numerical generation
of paths.
\end{abstract}

\goodbreak

\section{Introduction}

Differential equations have long been used to model the dynamics of
physical systems.  With the availability of computers, the tendency to
focus only on analysis of linear equations is being replaced by a
methodology that profits from a judicious mixture of numerical
generation of paths, bifurcation theory and asymptotic analysis.
However, when random perturbations (i.e. ``noise'') play an important
role, this new spirit is not so widespread.  One reason is that the
mathematical tools appropriate for describing stochastic paths are not
sufficiently well known.

 In Section 2 we introduce our notation, concentrating on
 the aspects of stochastic calculus  [1--6]
that may be
 unfamiliar to applied mathematicians.
In Section 3 we consider 
 the dynamic pitchfork bifurcation, which is of considerable 
interest in its own right [8--12],
and serves as an example of a system where noise of small 
magnitude has a disproportionate and simplifying effect 
\cite{landp,lythe4}.
In Section 4 we discuss some exact results for threshold
crossing problems that may be accessible to experiments
but have so far received little attention.
Our motivation comes from simple models for a single neuron's
membrane potential \cite{neuront,neuronms} and our intention
is to introduce new ways of extracting information from
numerical or experimental data.

A strong point of the stochastic approach is the close
 relationship between numerics and analysis. 
Throughout, we compare our calculations with numerical results,
 obtained using the algorithm described in Appendix 1. 

\subsection{Dynamic bifurcations}

When used to model
physical systems, differential equations
 have parameters representing external conditions
 or controllable inputs.
Bifurcation diagrams are used to depict the ultimate behaviour of 
their solutions as a function of  these parameters;
a critical value is a point on a boundary in the 
space of parameters separating areas with different ultimate behaviour.
 It is natural to
speak of a bifurcation ``taking place'' at a critical value.
Unfortunately, it is also common to speak of the qualitative 
behaviour of the system ``changing as a parameter passes through a
critical value'', although the paths of the
non-autonomous system of differential equations obtained by letting
a parameter become a function of time can differ
markedly from those of the equations with fixed parameters
 \cite{mande,benoit,arnold}.

Performing the analysis of Section 3, 
we have in mind the following situation. Looking for a symmetry-breaking
bifurcation, an experimentalist slowly changes 
a parameter until a qualitative change is seen. The change is
seen after the parameter value at which, mathematically speaking,
the bifurcation takes place;
the bifurcation is said to be delayed
 \cite{mande,benoit,arnold}.
Small random disturbances
 need to be considered explicitly to evaluate the 
magnitude of the delay. 
Previous work [13--17] on the dynamic pitchfork bifurcation
 has centered on the question: 
when do paths exceed a certain threshold distance from the 
former attractor? 
We calculate instead the last time at which this distance is zero.
The cases of small and large noise are treated, and
an approximate formula is presented,
with the appropriate limiting behaviors,
that also covers the intermediate region.

\subsection{Threshold crossing and neuron dynamics}

Animal brains are made up of very many neurons, that communicate by 
firing electrical signals [24--29].
A neuron `fires' when its 
membrane potential exceeds a threshold.  This membrane potential, the 
voltage difference between the interior of the cell and its 
surroundings, can be measured as a function of time.
  It is driven away from its rest state by 
many thousands of excitatory and inhibitory stimuli from other neurons 
and is sensibly modeled by a stochastic process.
We have in mind a situation where the SDE governing
the neuron is not known in advance \cite{neuronms}.
We examine the dependence of
 experimentally measurable quantities 
on the form of the SDE describing the paths until they reach a threshold,
concentrating on concepts that are a natural part of stochastic
calculus but not of traditional methods.
The hope is that measurement of these quantities will enable the
underlying dynamics to be deduced and lead to a greater 
understanding of single neurons and neural networks.

In Section 4, we consider
 the threshold crossing problem for
 a fairly general (non-linear) scalar SDE,
 but with constant threshold and noise intensity.
Those sufficiently well-acquainted with the probability
literature will not find new theoretical results; our
purpose is to present exact results
 that are amenable to experimental
and numerical verification, despite depending on subtle
properties of the paths on small scales, and to illustrate
by construction that exact results yield tractable
expressions even when the underlying SDE is nonlinear.
In Section 4.1 we split paths up into a series of excursions.
This provides, for example, a sensitive test
for asymmetries in the drift. In Section 4.2 we consider
the amount of time paths spend in intervals 
before reaching the threshold.
The function obtained in the limit where the intervals
shrink to points is the local time; for a stochastic
path, this is a continuous function.
Section 4.3 deals with the conditioning that comes from
considering only one part of a path and includes the calculation of
the density of the last time at zero.

\clearpage

\section{Stochastic calculus}

A stochastic differential equation
(SDE) is written in the following form: 
\begin{equation}
 \d\xt = f(\xt,t)\d t + \epsilon(\xt,t) \d{\bf W}_{t}.  
 \label{sde}
\end{equation}
Here \({\bf X}\) is a real-valued stochastic process; its
value at time \(t\) is a real-valued
random variable denoted by \(\xt\).
If \(\eps(x,t)\) is always zero then \eqref{sde} is just an 
ordinary differential equation;
 if \(\epsilon(x,t)\) is a real-valued function
independent of \(x\) then \eqref{sde} is a
differential equation with additive white noise.
The  Wiener process \({\bf W}\),
 also called standard Brownian motion,
 satisfies \eqref{sde} with \(f(x,t)=0\) and \(\eps(x,t)=1\).
 Its paths are continuous with probability \(1\).
For any \(t, \Delta t \ge0\),
 the random variable \({\bf W}_{t+\Delta t} - {\bf W}_t\) is Gaussian
 with mean \(0\) and variance \(\Delta t\), 
and successive increments are independent.

Random variables are in bold type in this work.
For any real-valued random variable \(\ba\), the probability that
\(\ba<x\) is an ordinary function of \(x\) denoted by
\(\pr{\ba<x} \). The  density of \(\ba\) (if it exists)
 is the derivative of this function with respect to \(x\):
\begin{equation}
R_{\ba}(x)={\d\over \d x}\pr{\ba<x}.
\label{densnot}
\end{equation}
This is sometimes put as:
 \(R_{\ba}(x)\d x\) is the probability that \(\ba\) lies in
 \((x,x+\d x)\). We maintain the notation \eqref{densnot},
with the random variable \(\ba\) as a subscript, except for
the density of \(\xt\), when we write simply \(\rxt\).

Solving an SDE on a computer consists of generating
 a set of values:
\(\{ {X}(t_i); i=1,\ldots , n \}\),
where \(0<t_1<t_2<\ldots < t_n=t\),  that 
approximate the values taken by one path of \( \xx \)
at the times \(t_i\). In the lowest order algorithm
\(X(t_{i+1})\) is generated from 
\(X(t_{i})\)  by adding a deterministic increment and a 
random one:
\begin{equation}
X(t_{i+1}) = X(t_i) + f(X(t_i),t_i)\Delta t + \epsilon(X(t_i),t_i)
 \sqrt{\Delta
t_i}\, n_i,
\label{alg}
\end{equation}
where \(\Delta t_i=t_{i+1}-t_{i}\) and
each \(n_i\) is independently generated from a Gaussian density 
with mean zero and variance 1.  
  Properties of the ensemble of paths, such as 
\(\mean{\xt}\), the mean value of \(\xt\), are
estimated by repeating this procedure as many times as necessary.

In the SDE \eqref{sde},
the coefficient \(f(x,t)\) is the mean displacement
or `drift':
if \(\xt=x\) then
\begin{equation}
f(x,t)=\lim_{\Delta t \to 0}\frac{1}{\Delta t}\mean{\xx_{t+\Delta t}-x}.
\label{driftdef}
\end{equation}
The second term on the right-hand-side of \eqref{sde} and \eqref{alg}
does not enter \eqref{driftdef} because it has mean zero. However, if
we consider the mean squared displacement then
\begin{equation}
\eps^2(x,t)=\lim_{\Delta t \to 0}\frac{1}{\Delta t}
\mean{(\xx_{t+\Delta t}-x)^2}.
\label{msdef}
\end{equation}
Further insight is gained by making the
following construction. Define
sets of times \(\lbrace t_i; i=0,1,\ldots,l\rbrace\) such that 
\(0=t_0<t_1<\ldots<t_l=t\) and \(t_{i+1}-t_{i} = t/l\).  Choose any 
path of \(\xx\) and let \begin{equation}[{\xx}]_t = 
\lim_{l\to\infty}\sum_{i=0}^{l-1}(\xx_{t_{i+1}}-\xx_{t_{i}})^2.
\label{qv}
\end{equation}
Then \([\xx]\), called the quadratic variation of \(\xx\),
satisfies 
\begin{equation}
\d[\xx]_t=\eps^2(\xt,t) \d t.
\label{simqv}
\end{equation}
(This is true of any path of \(\xx\) with probability 1.)
Note that if \(\eps\) is independent of \(\xt\) then
\eqref{simqv} is an ordinary differential equation.
The informal way to understand 
quadratic variation is to write
\begin{equation}
(\d \wt)^2=\d t.
\label{dws}
\end{equation}
Stochastic calculus  [1--6] is the calculus of (continuous but
non-differentiable) paths for which the
 limit \eqref{qv} is non-zero.

If \(\xx\) and \(\yy\) are stochastic processes
obeying SDEs of the form \eqref{sde}, then  the \ito\ integral,
\begin{equation}
\iint=\int_0^t \yy_s\d\xx_s,
\end{equation}
 is itself a stochastic process given by the following limit:
\begin{equation}
\iint= \lim_{\Delta t\to 0}\sum_{i=0}^{l-1}\yy_{t_i}
(\xx_{t_{i+1}}-\xx_{t_i}).
\label{itolimit}
\end{equation}
Its quadratic variation is
\begin{equation}
[{\bf I}]_t = \int_0^t \yy^2_s \d [\xx]_s.
\end{equation}

As suggested by \eqref{dws}, second order infinitesimals
are not always negligible in stochastic calculus. This is reflected in the
\ito\ formula, which is the chain rule of stochastic
calculus.  The SDE for \(f({\bf 
 X})\), where \(f\) is a \(C^2\) function, is related to that for 
 \({\bf X}\) by [1--6]
\begin{equation}
\d f(\xt) = f'(\xt)\d \xt + \ha f''(\xt)\d[{\bf X}]_t.
\label{itoform}
\end{equation}

An alternative definition of the stochastic integral is 
sometimes convenient,
 in which  \(\yy_{t_i}\) in \eqref{itolimit} is replaced
by \(\yy_{t_s}\), where \(t_s=\ha(t_i+t_{i+1})\).  If this alternative
definition, called the Stratonovich integral, is chosen then
 changes of variable can be performed without the
extra term that appears in \eqref{itoform}. (An extra term
appears instead in the numerical algorithm; in the Stratonovich
convention, \eqref{alg} no longer corresponds to the SDE \eqref{sde}.)

\eject

\section{The dynamic pitchfork bifurcation}

One of the most common transitions observed in nature is from a symmetric 
state to one of a set of states of lower symmetry
\cite{candh}.  Here we concentrate
on the example of a pitchfork bifurcation, described by the 
normal form \cite{gandh}
\begin{equation}
\dot x = gx-x^3.
\label{pitchfork}
\end{equation}
If the parameter \(g\) is less than \(0\) the solution \(x=0\) is stable;
if \(g>0\) it is unstable and the two states
\(x=\sqrt{g}\) and \(x=-\sqrt{g}\) are stable.
That is, for \(g>0\), any small perturbation away from \(x=0\) will
result in the system breaking the reflection symmetry
by selecting one of these states. In this work we examine what happens
if \(g\) is slowly increased through \(0\)
with white noise as the perturbation.

In a real experiment
or in a numerical search of the parameter space
of a set of differential equations,
 \(g\) will typically be a complicated function
of controllable parameters. Imagine that the experiment begins with \(g<0\)
and that the inputs are changed in such a way that it slowly increases.
In the normal form \eqref{pitchfork}, \(x\) has already been scaled so that
the coefficient of the cubic term is unity; we also rescale time so that
the starting value of \(g\) is \(-1\).
We thus solve the SDE
\begin{equation}
\d\xt = (\mu t\xt-\xt^3)\d t + \epsilon \d{\bf W}_t,\qquad \eps,\mu>0,
\quad t>t_0=-\frac1{\mu},
\label{pitchforkwn}
\end{equation}
and derive the statistics of the last time,  
\(\sss\),  at which \(\xt=0\). For convenience, 
we take initial condition \(\xtz=0\). However, we shall see that
 the presence of noise removes the dependence
on initial conditions for \(\mle<1\).

\begin{figure}
        \epsfysize=15cm
\epsfbox{dynpffig2.eps} 
\caption{ 
 {\em Dynamic pitchfork bifurcation.} 
Two paths of the SDE \protect{\eqref{pitchforkwn}} are shown.  In (a) 
noise is only important when \(\xt\) is near zero.
 The time of the jump away from zero is
calculated from the solution of the linearized equation.
Once one branch of the pitchfork is chosen, a further crossing of \(0\) is
improbable. In (b) we treat the problem as one of jumping between 
minima of a potential that is slowly deepening.
}
\label{trajs}
\end{figure}

Two paths of \eqref{pitchforkwn} are shown in Figure
 \ref{trajs}, where \(g=\mu t\).  The
first path lingers near \(\xt=0\) until \(g>\sqrt{\mu}\), jumps
away rather suddenly and does not return to \(0\). Although the path
looks smooth on the scale of the figure, the moment of this jump is in
fact controlled by the magnitude of the noise. The
characteristic value of \(g\) for the jump [13--16], 
\begin{equation}
\hat g = \sqrt{2\mle},
\label{ghat}
\end{equation}
is also found in the Hopf and transcritical bifurcations, and in the
analogous stochastic partial differential equations with a
time-dependent critical parameter \cite{lythe3}.  These bifurcations
share the property that a fixed point ceases to be an attractor, but
continues to exist, after the bifurcation \cite{nandb}.  On the other
hand, the dynamic saddle-node bifurcation, where a fixed point ceases
to exist at the critical point, has a characteristic delay that scales
as \(\mu^{{2\over 3}}\)\cite{jgrm}.

The result \eqref{ghat} is obtained by estimating
the value of \(g\) at which \(\xt\) is large enough that the 
linearized version of \eqref{pitchforkwn} is no longer valid.
  In Section 3.1, we
calculate the distribution of \(\sss\) using the same method and
show that it produces a good approximation for \(\eps\ll\sqrt{\mu}\).
  We also give an expression for the value of \(g\)
at which the distribution of \(\xt\) becomes bimodal.  In Section 3.2 we
consider the situation, illustrated in Figure 1(b), where paths
move back and forth between branches of the pitchfork numerous times
before settling down to one branch.  Here we derive the density of
\(\sss\) as an integral over the standard approximation for
 the rate of crossing between the two branches.
 This yields a good approximation
for the density of \(\sss\) when \(\eps\gg\sqrt{\mu}\). 
We close Section 3 with a
 formula that interpolates between the two limits.

\subsection{Linearized equation: small-noise case}

The solution of
the linearized version of \eqref{pitchforkwn}
(without the cubic term) is
\begin{equation}
\xt = \ee^{\ha\mu(t^2-t_0^2)} \left( \xtz + 
\epsilon\int_{t_0}^t \ee^{-\ha\mu(s^2-t_0^2)} \d \bs \right).
\label{linsoln}
\end{equation}
That is, \(\xt\) is a Gaussian random variable with mean
\begin{equation}
\mean{\xt}=\ee^{\ha\mu(t^2-t_0^2)}{\bf X}_{t_0},
\label{meanx}
\end{equation}
and variance
\begin{equation}
\mean{\xt^2}-\mean{\xt}^2=\var(t_0,t)=
\epsilon^2\ee^{\mu t^2}\int_{t_0}^t\ee^{-\mu s^2}\d s.
\label{sdx}
\end{equation}

Suppose that \(\xt=x\) at some time \(t\).
Let \(\tH\) be the first time after \(t\) at which
 \(\xx\) is \(0\),
with \(\tH = \infty\) if there is no such time. 
Note that, according to \eqref{linsoln},
\(\xt\to\pm\infty\) as \(t\to\infty\).
Because of the reflection symmetry,
if the path is at zero at some time then
it has an equal probability of
being positive or negative at a later time.
Thus the probability of never crossing
zero after time \(t\) is given by
\begin{equation}
\pr{\xx_\infty > 0|\xt=x} = \half \pr{\tH < \infty|\xt=x} +
\pr{\tH = \infty|\xt=x}.
\end{equation}
Since \( \pr{\tH = \infty|\xt=x} +\pr{\tH < \infty|\xt=x} =1\),
we find
\begin{equation}
\pr{\tH = \infty|\xt=x} =2\pr{\xx_\infty>0|\xt=x} - 1.
\label{thexp}
\end{equation}

We now evaluate the right-hand-side of \eqref{thexp}
using \eqref{meanx} and \eqref{sdx}. If \(\xt=x\) then,
for any \(T>t\), \(\xx_T\) is a Gaussian random variable
with mean m\((T)=\ee^{\ha\mu(T^2-t^2)}x\) and variance \(v(t,T)\).
Thus
\begin{eqnarray}
2\pr{\xx_T>0|\xt=x}-1
={\displaystyle 2\int_{0}^{{\rm m}(T)}}
(2\pi\var(t,T))^{-\ha}\exp\left(-{y^2\over 2\var(t,T)}\right)\d y.
\label{pxsgz}
\end{eqnarray}
Now making a change of variable in the exponent of \eqref{pxsgz} and
taking the limit \(T\to\infty\) gives
\begin{eqnarray}
\pr{\tH = \infty|\xt=x}
= {\rm erf}\left({x\over\epsilon}\ee^{-\ha\mu t^2}
\left(2\int_t^\infty\ee^{-\mu s^2}\d s\right)^{-\ha}\right)
\label{ptheinf}
\end{eqnarray}
where
\({\rm erf}(u)={2\over\sqrt{\pi}}\int_0^u \ee^{-y^2}\d y\).

The last time at zero is the random variable \(\sss\).
To calculate its density, we replace the initial condition  \(x\) 
by a Gaussian random variable with mean \(0\)
and variance given by \eqref{sdx}. 
Then \(H_1(t)=\pr{t>\sss}\) is obtained by  integrating 
\eqref{ptheinf} over the Gaussian density:
\begin{eqnarray}
H_1(t)&=&2{\displaystyle \int_{0}^{\infty}}
\pr{\tH = \infty|\xt=x}(2\pi\var(t))^{-\ha}
\exp\left(-{x^2\over 2\var(t)}\right)\d x\nonumber\\[5pt]
\label{ptgs}
&=&{\displaystyle {1\over\sqrt{\pi}}\int_0^\infty}
\ee^{-u^2}{\rm erf}\left( u \left({
\int_t^\infty \exp(-\mu s^2)\d s\over
\int_{t_0}^t \exp(-\mu s^2)\d s}\right)^{\ha}\right)
 \d u\\
&=&{\displaystyle {2\over\pi}}
\rm{atan}\left(\left({{
\int_t^\infty \exp(-\mu s^2)\d s}\over
\int_{t_0}^t \exp(-\mu s^2)\d s}\right)^{\ha}\right).\nonumber
\end{eqnarray}
The density of the last time, \(\sss\), 
at which \(\xt\) passes through \(0\) is 
\begin{equation}
R_{\sss}(t) = {\d\over \d t}H_1(t)=
{1\over \pi} {\exp(-\mu t^2)
\over\left(
\int_{t_0}^t \exp(-\mu s^2)\d s
\int_t^\infty \exp(-\mu s^2)\d s\right)^{\ha}
}.
\label{linlast}
\end{equation}
For \(\mu\to 0\), this density is
symmetric about \(t=0\) with width proportional to
\(\frac{1}{\surd\mu}\).
 The above calculations do not include
the effect of the cubic term in \eqref{pitchforkwn}.
Because this term  always pushes paths towards
\(\xt=0\), the function \(H_1(t)\) from \eqref{ptgs} is an upper bound on
 the probability of no crossing of zero after time \(t\).

Self-consistency of the approximation \eqref{ptgs} demands that
\(\xt^3\ll g\xt\) for \(t<\frac{1}{\surd\mu}\).
From \eqref{sdx}, \(\xt\sim\eps\) 
for \(t<\frac1{\surd\mu}\). Thus we require \(\eps\ll\sqrt{\mu}\).
The calculations of this Section are quantitatively accurate when,
in addition,  the probability of a crossing of zero 
for \(t\gg\frac1{\sqrt{\mu}}\)
is negligible, corresponding to the
situation shown in Figure 1(a). This assumption will be
 examined in the next subsection;
it is also true for \(\eps\ll\sqrt{\mu}\).

We have used a zero initial condition 
in \eqref{pitchforkwn} because it is convenient
to have \(\mean{\xt}=0\).
 This initial condition  corresponds to the case where
the system is already close to its equilibrium before the sweep is started.
With a non-zero initial condition,
our calculations remain valid if \(\mean{\xt^2}\gg\mean{\xt}^2\)
 for \(t>0\).  Comparison of the magnitudes of \eqref{meanx}
(proportional to \(\xtz\))
and \eqref{sdx} (proportional to \(\eps\))
shows that a non-zero initial condition 
is unimportant unless \(\xtz>\eps\exp({1\over2\mu})\).
 This insensitivity to initial condition is characteristic
of the simplification produced by noise in problems where
 paths sweep repeatedly past an invariant manifold \cite{landp,lythe4}.

The results of this Section are changed little if the
noise is `colored' (i.e. if \(\wt\) in \eqref{pitchforkwn}
is replaced by a stochastic process whose increments are correlated)
\cite{lythe5}.
If a deterministic perturbation is used  instead of noise, the effect
is proportional to the mean of the perturbation. Multiplicative
noise is less important than additive noise \cite{lythe5,lythe3}.

\subsubsection{Transition to bimodal density}

To estimate when nonlinear terms become important in a dynamic
bifurcation, one can calculate from the linearized version
the time when \(\xt\) exceeds some threshold [13--16].
Here, we calculate instead, from the
nonlinear equation with \(\epsilon\ll\sqrt{\mu}\ll 1\),
the time when the distribution of \(\xt\)
changes from unimodal (one maximum at \(0\)) to bimodal.
We do this by considering the evolution
of an ensemble of paths from a
Gaussian distribution of initial conditions
into the latter part of the path where
\(|\xt|\to\sqrt{g}\) from below.
Effectively, noise provides a random initial condition for
the subsequent, deterministic, evolution.

The solution of the deterministic equation,
\begin{equation}
\dot x = \mu t x -x^3,
\end{equation}
with initial condition \(x=x_1\) at \(t=t_1\) is:
\begin{equation}
 x= \frac{x_1}{|x_1|} \ee^{\ha\mu( t^2-t_1^2)}
\left(x_1^{-2}+2\int_{t_1}^t\ee^{\mu(s^2-t_1^2)}\d s\right)^{\ha}.
\label{noncub}
\end{equation}
Now, instead of a constant as initial condition, we 
take the following Gaussian random variable:
\begin{equation}
 {\bf X}_{t_1}=\limvar(t_1){\bf n},
 \qquad{\rm where}\qquad \limvar(t) =
\epsilon\left({\pi\over\mu}\right)^{{1\over4}}\ee^{\ha\mu t^2}
\label{gic}
\end{equation}
is the large-\(t\) limit of \(\var(t)\) \eqref{sdx},
and \({\bf n}\) is a Gaussian random
variable with mean \(0\) and variance \(1\).
 We use
\(2\int_{t_1}^t 
\exp(\mu s^2)\d s \simeq \exp(\mu t^2)(\mu t)^{-1}\)
 and obtain
\begin{equation}
\xt^2 = \left(
{1\over \limvar^2(t) {\bf n}^2}+{1\over \mu t}\right)^{-1}.
\end{equation}
The density of \(\xt\) is then given, with \(t\) as a parameter, by
\begin{equation}
\rxt = \sqrt{{2\over\pi}}\limvar(t)\exp\left(
\frac{-x^2}{ 2\limvar(t)^2 (1- x^2/\mu t)}\right)
\left(1-{x^2\over \mu t}\right)^{-{3\over 2}};
\label{xtdens}
\end{equation}
 it has a maximum at a non-zero value of \(x\) if
\(\limvar^2(t)>\frac13\mu t\).
The time when the density changes to  bimodal therefore satisfies
\begin{equation}
(\mu t)^2-\mu\log {\mu t\over3} = 2\mu|\log\epsilon| 
+\frac{\mu}{2}\log{\pi\over\mu}.
\label{sqrtg}
\end{equation}
For small \(\mu\), we recover the result 
\((\mu t)^2 = 2\mu|\log\epsilon|\) that comes from the analysis
of the linearized system with a fixed threshold \cite{tsm,smm,sha,landp}.

\subsection{Effect of the cubic term: large-noise case}

The probability that there is no crossing of
zero after time \(t\) is a positive increasing function 
of time, \(H(t)\).
We can therefore make the decomposition
\begin{equation}
\frac{\d}{\d t}H(t) = r(t)H(t).
\label{ddtht}
\end{equation}
So
\begin{equation}
H(t)=
\exp\left(-\int_t^\infty r(s)\d s\right).
\label{htexp}
\end{equation}
The function \(r(t)\) is everywhere positive
and can be interpreted as the 
probability per unit time of a crossing of zero.
In Section 3.1 we obtained an upper bound on \(H(t)\)
from the linearized equation.
A lower bound on \(H(t)\) can be obtained as follows.
Calculate the probability per unit time of
a crossing of zero from \eqref{pitchforkwn}
with \(\mu t\) taken as a fixed parameter.
Denote this function by \(r_2(t)\).
Because the slow increase of \(g\) makes crossings of zero rarer
as \(t\) increases, \(r_2(t)\) is an upper bound
on the rate of zero crossings.
The function  \(H_2(t)\), calculated from \eqref{htexp},
is therefore a lower bound on the exact result \(H(t)\).
For direct comparison
with numerics, \(H_2(t)\) is obtained by integrating
 \(r_2(t)\) over the density
of \(\xt\); the result is an unwieldy expression.
Here we restrict ourselves to the case \(g>\eps\) and
examine the validity of the compact expression that results.

The expression \eqref{ddtht} is most useful and intuitive for
\(g>\eps\), when paths spend most of their
time near \(\pm\sqrt{g}\).
Then \(r_2(t)\) is well approximated by
the inverse of the mean passage time
from \(\xt=\pm\sqrt{g}\) to \(\xt=0\)
\cite{Gardiner,nandb}:
\begin{equation}
r_2(t)={\sqrt{2}\over\pi} g\, \exp\left(-{g^2\over 2\epsilon^2}\right).
\label{r2}
\end{equation}
Using \eqref{r2} in \eqref{ddtht} with \(g=\mu t\) gives
\begin{eqnarray}
H_2(t)
=\displaystyle{\sqrt{2}\over \pi}\exp\left(
-{\sqrt{2}\over \pi}{\epsilon^2\over\mu}
\exp\left(-{g^2\over 2\epsilon^2}\right)\right).
\label{h2texp}
\end{eqnarray}
The expression for the density of \(\sss\) 
corresponding to \eqref{h2texp} is
\begin{equation}
R_{{\bf s}}(t)= \frac{\d}{\d t}H_2(t)=
{\sqrt{2}\over \pi}g\,\exp\left(-{g^2\over 2\epsilon^2}\right)
\exp\left(
-{\sqrt{2}\over \pi}{\epsilon^2\over\mu}
\exp\left(-{g^2\over 2\epsilon^2}\right)
\right).
\label{lct}
\end{equation}
As \(\mu\to 0\),
the mean value of \({\bf s}\) is \cite{landp}
\begin{equation}
\mean{{\bf s}} = \frac1\mu\left(g^* + \gamma\frac{\eps^2}{g^*}\right)
\qquad{\rm where}\qquad
g^*=\eps\left(2\log\left(\frac{\sqrt{2}\eps^2}{\pi\mu}\right)\right)^{\ha} 
\label{gstar}
\end{equation}
and Euler's constant \(\gamma=0.57721\ldots\).
In the same limit, the standard deviation is
\begin{equation}
\left(\mean{{\bf s}^2}-\mean{{\bf s}}^2\right)^{\ha}=
\frac{\eps \pi}{\mu\sqrt{12}}
\left(\log\frac{\sqrt{2}\eps^2}{\pi\mu}\right)^{-\ha}.
\label{sds}
\end{equation}
The approximations \eqref{r2}-\eqref{sds} are valid 
 if \(g^*\gg\eps\), which is true if \(\eps>\sqrt{\mu}\).
Note also that \(H_2(t)\simeq 1\) for \(g\ge\sqrt{\mu}\)
if \(\eps < \sqrt{\mu}\), as required for validity
of the calculations of Section 3.1.

The decomposition \eqref{ddtht} also provides a natural way to
unify the above approximation with that of Section 3.1. Let
\(r(t)=r_1(t)+r_2(t)\) where 
\(r_1(t)=H_1(t)^{-1}\frac{\d}{\d t}H_1(t)\),
with \(H_1(t)\) given by \eqref{ptgs}.
 Then \(H(t)\) can be obtained from \eqref{htexp}.
This yields the following
 approximation to \(R_{{\bf s}}(t)\)
that has \eqref{lct} and \eqref{linlast}
 in the large-noise and small-noise regimes:
\begin{eqnarray}
R_{{\bf s}}(t)
&=&(r_1+r_2)H_1H_2\nonumber\\
&=&H_2(t){\d\over\d t}H_1(t) + H_1(t){\d\over\d t}H_2(t).
\label{combf}
\end{eqnarray}
In Figure \ref{formvsnum}, we used
\begin{eqnarray}
r_2(t)  = \left\{ \begin{array}{ll}
 {\sqrt{2}\over\pi} g\, 
\exp\left({-{g^2\over 2\epsilon^2}}\right)\ & g\ge\epsilon,\\
 {\sqrt{2}\over\pi} \epsilon\, \exp\left(-{1\over 
2}\right)\ & g<\epsilon.\end{array}\right.
\label{r12}
\end{eqnarray}

\begin{figure}
\epsfbox{dynpffig3.eps}
\caption{ 
{\em Last crossing of zero: numerical results vs formula}.
The dots are the histogram compiled from numerical generation 
of 10000 paths of \protect{\eqref{pitchforkwn}}
and the solid lines are the uniform approximation
described in Section 3.3.
Figure (a) exhibits the small-noise case
(\(\epsilon<\surd\mu\)) and (b) the large-noise case.
Note \(g=\mu s\).
}
\label{formvsnum}
\end{figure}

\clearpage

\section{Exact results on threshold crossings}

In this Section we consider processes 
obeying SDEs of the following type:
\begin{equation}
 \d\xt = f(\xt)\d t + \d\bt
 \label{neurongensde} 
\end{equation}
with initial condition \(\xz=0\).
 Paths are followed until the first time, \(\tb\), that
 they reach a threshold \( b>0 \).
If \(f(x)\) is not known in advance,
what can be learned from the paths of \(\xt\) for \(t<\tb\)?
The process
 \(\xt\) models, for example,
 the membrane potential of a neuron \cite{neuront,tuck2}
or the path of a diffusing particle \cite{jandr}.
To produce \eqref{neurongensde} from the more general
 SDE \eqref{sde}, we assume that \(\eps\) is constant.
We can then scale time so that
the coefficient of the second term in \eqref{sde} is \(1\).
We also primarily  consider the case where the drift, \(f(x)\),
is towards \(x=0\).

We do not calculate two of the quantities traditionally considered by
theorists: the density of \(\xt\) and that of \(\tb\)
\cite{neuronk,candg}.  We calculate instead quantities that are
measurable by an experimentalist who can record individual paths, but
have so far received little attention outside of probability theory.
All quantities calculated are derived from the following function:
\begin{eqnarray}
S(x)=\left\{ \begin{array}{ll}
\displaystyle\int_{0}^{x}\exp\left(2U(y)\right)\d y & x\ge0,\\[15pt]
\displaystyle\int_{x}^{0}\exp\left(2U(y)\right)\d y & x<0,\end{array}\right.
\label{sxdef}
\end{eqnarray}
where
\begin{eqnarray}
U(x)=-\int_{0}^{x}f(y)\d y.
\label{udef}
\end{eqnarray}
The function \(S(x)\) is generated on a computer by
solving the ordinary differential equation 
\begin{eqnarray}
\begin{array}{llll}
S'(x)&=&\exp(2U(x)) \qquad &x>0,\cr
S(0)&=&0,\\
S'(x)&=&-\exp(2U(x)) \qquad &x<0.
\end{array}
\end{eqnarray}
 Further, \(S(x)\)
has the following simple interpretation.
Given \(\xt=x\), \(0\le x\le b\), at some time,
 the probability that
\(\xx\) reaches \(b\) before \(0\) is \(\frac{S(x)}{S(b)}\).
An advantage for performing calculations
is that \(S(x)\)
 is time-independent. This is a consequence of the
`memoryless' property of paths of the SDE:
the future of \(\xt\) depends only on the value of \(\xt\),
not on the path before this time.
In the traditional approach,
based on the Fokker--Planck equation, measurable quantities
are derived from the density of \(\xt\), which
 is time-dependent because of the flux of probability through
the threshold.

A path of a stochastic process obeying the SDE \eqref{neurongensde}
can be thought of as being made up of a 
series of excursions from zero.  
Each time \(\xt\) returns to \(0\) the process is
considered as starting afresh, and each excursion 
treated as an independent event [2--5].
 In Section 4.1  we derive
the statistics of the excursions in terms of the maximum distance
from the origin during the excursion.

In Section 4.2 we use the concept of local time, which measures
the occupation density of a path \cite{randw,kands,randy,dandj}.
 Given a path of a stochastic process satisfying \eqref{neurongensde},
 the local time at \(a\)
 is a stochastic process 
 whose value at time \(t\) is given by
\begin{equation}
{\bf L}_t(a) 
= \lim_{\Delta a\to 0}\frac{1}{2\Delta a}\int_0^t 
\ind{\xs\in(a-\Delta a,a+\Delta a)} \d s,
\label{ladef}
\end{equation}
where \({\cal I}\{\Theta\}=1\) if \(\Theta\) is true and is zero otherwise.
Note that for fixed \(a\), 
\({\bf L}_t(a) \) increases at times 
when \(\xt=a\) and is constant otherwise. 

Local time is the  limit of functions that can be constructed from
a series of values of \(\xt\) that could be available
to an experimentalist. Suppose
the amount of time spent in boxes \((a-\Delta a,a+\Delta a)\)
before \(\tb\) is recorded.
The function obtained in the limit \(\Delta a\to 0\)
is \(\ltba\).
 In Section 4.2 we consider two ways of comparing
numerical or theoretical data with \(\ltba\): as an average over
many paths and for one path at a time.

In Section 4.3 we consider the portion of the path
of \(\xx\) after \(\ssb\), the last time that the path 
passes through zero before \(\tb\).
This is convenient, both numerically and experimentally,
when \(\tb\) is large: rather than generating or recording
the whole path, one can concentrate on only the last portion.
Further, it is one case where calculations based on a constant threshold
will be a good approximation to the case where
the threshold approaches a constant value for large \(t\).

\begin{figure}
\epsfbox{dynpffig1.eps}
\caption{ 
{\em Threshold crossing.}
 The first time a path attains a level \(b\) is
a random variable, denoted by \(\tb\).
 Also indicated is \(\ssb\), the last time
before \(\tb\) at which the path is at \(0\). 
The path shown is a path of the Wiener process.
}
\label{fpt}
\end{figure}

\clearpage

\subsection{Excursions}

An excursion is a portion of a path between successive crossings of \(0\).
A path can be thought of as consisting of a series of
excursions occurring one after the other, with properties
chosen according to a certain probability.
We construct this probability as follows. For \(x>0\),
let \(N(x)\) be the number of excursions before \(\ssb\)
(see Figure \ref{fpt}) during which the maximum of \(\xx\) is
greater than \( x\); for \(x<0\),
let \(N(x)\) be the corresponding number of excursions with 
minimum less than \( x\).

Suppose that at some time \(t\), \(\xt=x>0\). Then the probability that
\(\xx\) reaches \(b\) before \(0\) is \(\frac{S(x)}{S(b)}\).
Thus
\begin{eqnarray}
\pr{N(x)=n}=\frac{S(x)}{S(b)}\left(1-\frac{S(x)}{S(b)}\right)^n
\qquad x>0.
\label{nxp}
\end{eqnarray}
Now consider \(x<0\) and
suppose that at some time \(t\), \(\xt=0\). Then the probability that
\(\xx\) reaches \(x\) before \(b\) is 
 \(\frac{S(b)}{S(x)+S(b)}\).
Thus
\begin{eqnarray}
\pr{N(x)=n}=\frac{S(x)}{S(x)+S(b)}\left(\frac{S(b)}{S(x)+S(b)}\right)^n
\qquad x<0.
\label{nxn}
\end{eqnarray}
From \eqref{nxp} and \eqref{nxn} we find the following exact result
for the mean of \(N(x)\):
\begin{eqnarray}
\mean{N(x)}=\left\{\begin{array}{lll}
 &\frac{S(b)}{S(x)}
  & x< 0,\\
&\infty&x=0\\
&\frac{S(b)}{S(x)}-1  & x>0.\end{array}\right. 
\label{nx}
\end{eqnarray}
We find good agreement between the exact result \eqref{nx} and 
 averages over paths of generated with a finite time-step
 (see Figure \ref{excur}).
Note that, if the drift \(f(x)\) is symmetric, then \(\mean{N(-b)}=1\).
This is a sensitive test for asymmetries in \(f(x)\).
We have not counted the portion of the path after \(\ssb\)
 as an excursion. We return to this
in Section 4.3.

\begin{figure}
\epsfbox{dynpfx.eps}
\caption{
{\em Pre-threshold excursions.}
Here \(\mean{N(x)}\) is the average number of excursions
with maximum greater than \(|x|\) before \(\tb\)  (\(b=2\)).
The curves, produced 
by averaging over 100 numerically-produced paths
with \(\Delta t=10^{-4}\),
are in excellent agreement with the exact result \protect{\eqref{nx}}.
}
\label{excur}
\end{figure}

\clearpage

\subsection{Occupation density}

Another way to extract information from a path
followed until it reaches a threshold is to construct
the occupation density:
the amount of time spent in boxes \((a-\Delta a,a+\Delta a)\)
before \(\tb\).
The function obtained in the limit \(\Delta a\to 0\)
is \(\ltba\), the local time for a path run until it first reaches the threshold \(b\).
For convenience, in this Section we denote \(\ltba\) simply by \(\la\).

We first consider the local time, averaged over many paths.
Given \(\xz=x\), \(c<x<b\), let \(l(x)\) be the mean value
of the local time at \(a\) before the first time that \(\xt\)
exits the interval \((c,b)\).
Using the delta-function notation to rewrite \eqref{ladef},
\begin{equation}
\la
= \int_0^{\tb} \delta(\xs-a) \d [{\bf X}]_s,
\end{equation}
we find (see Appendix 2)  that \(l(x)\) satisfies:
\begin{eqnarray}
-\delta(a)=f(x)l'(x)+\ha l''(x), \qquad l(c)=l(b)=0.
\label{lampde}
\end{eqnarray}
With \(x=0\) and \(c=-\infty\) we obtain \(\mean{\la}\)
from the solution of \eqref{lampde}:
\begin{eqnarray}
\mean{\la}=\left\{\begin{array}{lll} &2\exp(-2U(a))(S(b)-S(a))
  & a\ge 0,\\
&2\exp(-2U(a))S(b) & a<0.\end{array}\right. 
\label{meanlta}
\end{eqnarray}
This average is displayed for several choices of \(f(x)\)
in Figure \ref{lamfig}.

\begin{figure}
\epsfbox{dynpflam.eps}
\caption{ {\em Average occupation density}.  
The curves are \protect\eqref{meanlta}, the analytical result
for the average (over many paths) of the amount of time spent 
at \(a\) before the first time that a path, started at \(0\),
reaches \(b=2\). 
}
\label{lamfig}
\end{figure}

The average \eqref{meanlta} is an average over realizations
and thus  is useful  when
many paths have been recorded. 
Quantitative comparison can also be performed
for just a few paths, exploiting the fact that 
an occupation density can  be constructed from just one path.
  A remarkable fact
is illustrated in Figure \ref {lafig}:
\(\la\) is a continuous function of \(a\) when \(\eps\ne 0\).
In fact, \(\la\) can be considered as
a stochastic process {\em indexed by the space variable} \(a\).
In Appendix 3 we show that
 \(\la\) satisfies the non-autonomous SDE \cite{nrw}
\begin{eqnarray}
\d{\bf L}_z = \left\{\begin{array}{lll}
2 {\bf L}_z^{\half} \d \wz - 2(f(b-z){\bf L}_z-1) \d z &z\le b,\\
2 {\bf L}_z^{\half} \d \wz - 2f(b-z){\bf L}_z \d z &z>b,\end{array}\right.
\label{spacesde}
\end{eqnarray}
where \(z=b-a\).
Traces  mimicking those of Figure \ref{lafig} can therefore
be directly generated by solving \eqref{spacesde} with the initial
condition \({\bf L}_0=0\). The drift of
the SDE \eqref{spacesde} ensures that \({\bf L}_z\)
remains positive for \(z<b\) (\(a>0\));
a path is followed until \({\bf L}_z=0\) for some \(z>b\) (\(a<0\)).

\begin{figure}
\epsfbox{dynpfla.eps}
\caption{
 {\em Occupation density}.  
Each curve was obtained by following one path 
until it reached \(b=2\). We used \(\Delta a=0.01\) and \(\Delta t=10^{-5}\).
Note the discontinuities in \(\la\) for the deterministic path. 
}
\label{lafig}
\end{figure}

For \(a>0\), the density of the random variable \(\la\) is
\begin{equation}
R_{\la}(y)=\mean{\la}^{-\ha}
\exp\left(-\frac{y}{\mean{\la}}\right),
\end{equation}
with \(\mean{\la}\) given by \eqref{meanlta}.
The corresponding result for \(a<0\) is
\begin{eqnarray}
\pr{\la=0}&=&\frac{S(a)}{S(a)+S(b)},\\
R_{\la}(y)&=&\frac{S(b)}{S(a)+S(b)}\frac1{q(a)}\exp(-\frac{y}{q(a)})
\qquad y>0,
\label{ladensn}
\end{eqnarray}
where \(q(a)=2\exp(-2U(a))\left(S(a)+S(b)\right)\).

\clearpage

\subsection{Last zero crossing}

When \(\tb\) is large, it may be
 impractical experimentally to record the whole path
until \(\tb\). Numerically, it also becomes
time-consuming to generate enough paths to
evaluate averages. One way around this is to concentrate
on the portion of the path between \(\ssb\),
 the last time that the path 
passes through zero, and \(\tb\). Numerically,
it is possible to generate this portion of the path directly.

Generating one portion of a path conditioned on desired
properties is equivalent to selecting a subset of 
possible paths (here, paths that do not return to \(0\)).
This is equivalent to changing the drift of the SDE.
Define \(\yy\)  to be the conditioned process and
suppose \(\xt=x\) at some \(t\). Then the probability
that \(t>\ssb\) is \(h(x)=\frac{S(x)}{S(b)}\).
That is, the probability that a path of \(\xx\) with \(\xt=x\)
is also a path of \(\yy\) is \(h(x)\).
Let \(\Delta \xx=\xx_{t+\Delta t}-\xt\). Then
the drift of the SDE for \(\yy\) is
\begin{eqnarray}
\lim_{\Delta t\to 0}\frac1{\Delta t}
\mean{\yy_{t+\Delta t}-x} 
&=&\lim_{\Delta t\to 0} \frac1{\Delta t}
\frac{\mean{ h(x+\Delta \xx)\Delta\xx}}{\mean{h(x+\Delta \xx)}}\nonumber\\[5pt]
&=&\lim_{\Delta t\to 0}\frac1{\Delta t} 
\frac{\mean{\left(h(x)+h'(x)\Delta\xx+\ldots\right)\Delta \xx}}{h(x)}\\[5pt]
&=& 
\lim_{\Delta t \to 0}\frac{1}{\Delta t}\mean{\xx_{t+\Delta t}-x}
 + {S'(x)\over S(x)}.
\nonumber
\end{eqnarray}
Because we are choosing a subset of
paths with non-zero probability, the 
rate of increase of quadratic variation 
(which is the same for each path) is not affected
by the conditioning.
 The SDE for \(\ys\) is thus
\begin{equation}
\d\ys = \tilde f(\ys) \d s + \d\ws 
\qquad{\rm where}\qquad
\tilde f(y) = f(y) +  \frac{S'(y)}{S(y)}.
\label{condsde}
\end{equation}
The transformed drift \(\tilde f(x)\) is singular at \(x=0\),
which has the required effect of preventing paths
of \(\yy\) from returning to \(0\). 
(Note the
following relationship between \(f\) and \(\tilde f\):
\begin{equation}
\tilde f'(x) = f'(x)-\tilde f^2(x)+f^2(x).
\label{fftildede}
\end{equation}
The boundary condition for \eqref{fftildede} is 
 \(x\tilde f(x) \rightarrow 1\) as \(x \rightarrow 0\).)

From the transformed drift, quantities such as 
 \(\tilde U'(y)=-\tilde f(y)\) can be defined and
calculated as described in this section. For example,
the mean value of \(\tb-\ssb\) is the mean first passage time
from \(\yy=0\) to \(\yy=b\), which is given by
(Appendix 2):
\begin{eqnarray}
\mean{\tb-\ssb}
=& \displaystyle\lim_{x\to 0} 2\displaystyle\int^b_{x}
\exp{\left({2}\tilde U(y)\right)}
\int_{0}^y \exp{\left(-{2}\tilde U(z)\right)}\d z\, \d y\\[10pt]
=&\displaystyle\lim_{x\to 0} 2\displaystyle\int^b_{x}
\frac1{S^2(y)}\exp{\left({2}U(y)\right)}
\int_{0}^y S^2(z)\exp{\left(-{2} U(z)\right)}\d z\, \d y.\nonumber
        \label{ltz}
\end{eqnarray}
In Figure \ref{passt} this mean is displayed as a function
of the threshold \(b\) for several examples of \(f(x)\).

\begin{figure}
\epsfbox{dynpflz.eps}
\caption{{\em Last zero crossing}.
The curves are $\tb-\ssb$ versus the height of the boundary $b$
 for various choices of the drift $f(x)$.
They are produced by integrating 
the expression \protect\eqref{ltz}.
}
\label{passt}
\end{figure}

\clearpage

\section{Conclusion}

If a symmetry-breaking bifurcation is searched for by making a 
parameter a function of
time, the delay produced means that 
it is not obvious how to define
or measure its position using standard ideas.
The presence of noise simplifies the picture
by removing the dependence on initial conditions
in the limit of small sweep rate, but means that the
usual criteria for determining the position of the bifurcation,
 based on deterministic differential equations,
are no longer appropriate \cite{nandb}. 
We study the dynamic pitchfork bifurcation and
choose a simple concept:
the last time, \({\bf s}\), that a path is at zero.
The density of \({\bf s}\) is calculated as a function
of the noise level, \(\eps\), and the rate of change
of the parameter, \(\mu\).
 If \(\epsilon\ll\sqrt{\mu}\), there is a neat
 separation between the noise-dominated
and nonlinear parts of the path. The mean
of \({\bf s}\) in this case is the point of loss
of stability as calculated by static theory and
 the standard deviation of \({\bf s}\) is
proportional to \(\mu^{-\ha}\).  
An expression is also calculated for the time when the density 
of \(\xt\) becomes two-humped. Here the  result is
 consistent with those obtained by setting a threshold on \(\xt\):
the characteristic
delay of the bifurcation is \(\sqrt{2\mle}\).
In the second case, \(\epsilon\gg\sqrt{\mu}\), 
the analysis is analogous to that of a two-state system.
The mean value of \({\bf s}\) is a slowly increasing function of \(\mu\) 
 and \(\eps\),
and the standard deviation of \(\sss\) is proportional
to \(\eps\).
Numerical results are compared with a uniform approximation
that captures these two cases as limits.

A threshold crossing problem is the ideal setting for
stochastic methods. In Section 4, we
consider some exact results and compare them with
numerical data. Our aim is to
 calculate quantities that an experimentalist,
confronted with a series of paths like that of Figure \ref{fpt},
can use to deduce the underlying dynamics. 
The particular motivation comes from the modeling
of neurons that fire when their membrane
potential exceeds a threshold.
Results are expressed
in terms of the function \(S(x)\) \eqref{sxdef},
which is easily generated on a computer. 

\section*{Acknowledgements} 
We are grateful for the encouragement of Mike Proctor,
Nigel Weiss and Jer\^ome Losson.

\appendix

\section*{Appendices} 

\subsection*{1. Numerical methods}

As suggested by
the SDE notation \eqref{sde}, the simplest
algorithm for solving an SDE is the following.
At each step add two increments to \(\xt\): a drift given by
\(f(\xt,t)\Delta t\), where \(\Delta t\) is the timestep, and another
that is a random variable proportional to \(\epsilon\sqrt{\Delta t}\).
The stochastic version of the Euler method
for constructing an approximation
\(X_t\) to  \(\xt\) is:
\begin{equation}
X_{t+\Delta t} = X_t + f(X_t,t)\Delta t + \epsilon\,(X_t,t) \sqrt{\Delta t}\,n
\label{steu}
\end{equation}
where \(n\) is a Gaussian random variable with mean zero and variance 1.
The method that was used to generate the figures in this work
is analogous to the second-order Runge-Kutta
 method \cite{gsh}:
\begin{eqnarray}
&X_{t+\Delta t} - X_t =\cr &\ha\left( 
f(X_t,t) + f\left(X_t+f(X_t,t)\Delta t + \epsilon \sqrt{\Delta t}\,n
,t+\Delta t\right)\right)
\Delta t +\epsilon \sqrt{\Delta t}\,n.
\label{srkt}
\end{eqnarray}
Note that only one Gaussian random variable is generated at each time step.
For the algorithm \eqref{srkt}
there exist positive constants \(\kappa\) and \(\delta\) such that, 
if \(\eps\) is constant and if the
step size \(\Delta t\) is less than \(\delta\), then 
\(
\left|\mean{h(X_t)}-\mean{h(\xt)}\right|\le \kappa(\Delta t)^2
\)
for any polynomial function \(h\).
Higher order algorithms exist \cite{kandp}.

\eject

\subsection*{2. Derivation of PDEs from the \ito\ formula}

Consider the SDE
\begin{equation}
\d \xt = f(\xt)\,\d t + \epsilon\, \d \wt, \qquad {\bf X}_0=x,
\qquad a<x<b,
\label{tsde}
\end{equation}
and define \(T(x)\) to be the mean value of the first time
that \(\xt=b\). 
Since \(T(x)\) is an ordinary function of the initial condition \(x\),
we can consider the evolution of \(T(\xt)\) along a path.
The SDE for \(T(\xt)\) is, using the \ito\ formula \eqref{itoform}:
\begin{equation}
\d T(\xt) = \tx(\xt) \d X_{t} + \ha\eps^2\txx(\xt) \d [{\bf X}]_t.
\label{dbigt}
\end{equation}

We know, 
 from the definition
of \(T(x)\), that
\begin{equation}
\mean{\d T(\xt)}=-\d t.
\label{meandtx}
\end{equation} 
From \eqref{simqv}, the quadratic variation for the process \(\xt\) satisfies
 \(\d[\xt]=\epsilon^2\,\d t\) and, from \eqref{driftdef},
\(\mean{\d \xt} = f(\xt)\d t\).
Thus taking the mean value of both sides of
\eqref{dbigt} gives \cite{Gardiner,kandt}
\begin{equation}
\ha\epsilon^{2} \txx + f(x) \tx =  -1
,\qquad T(a)=T(b)=0.
\label{fptpde}
\end{equation}

In Section 4, we also consider the quantities
\(h(x)\), the probability that \(\xx\) reaches \(b\) before \(a\),
and \(l(x)\), the mean value of \(\ltba\). The appropriate PDEs
are derived as above, using 
\(\mean{\d h(\xt)}=0\) and \(\mean{\d l(\xt)}=-\delta(a)\d t\)
 instead of \eqref{meandtx}.

\eject

\subsection*{3. SDE for local time in space variable}

As in Section 4.2, \(\la\) denotes the local time at \(a\) \eqref{ladef},
evaluated at \(t=\tb\).
We consider \(\la\) as a stochastic process indexed by \(a\)
and derive the corresponding SDE.
By differentiating \eqref{meanlta},
the drift of the SDE for \(\la\) is:
\begin{equation}
\frac{\d}{\d a}\mean{\la}=\left\{\begin{array}{lll}
2(f(a)\la-1)&x\ge0,\\
2f(a)\la&x<0.
\end{array}\right.
\end{equation}

 To find the quadratic variation of \(\la\), 
we begin by using \ito's formula to derive an SDE
for the function \(u(\xt)\) where \(\xt\) satisfies \eqref{neurongensde} 
and 
\begin{eqnarray}
u(x)=&\left\{\begin{array}{ll}x \qquad x\ge a,\\
a \qquad x<a.\end{array}\right. 
\end{eqnarray}
Note that \(u''(x)=\delta(x-a)\).
 The SDE for \( u(\xt)\) is
\begin{equation}
\d u(\xt) = \ind{\xt>a}\d \xt + \half\delta(\xt-a)\d t.
\label{usde}
\end{equation}
Integrating \eqref{usde} yields 
\begin{equation}
 u(\xt) - u(\xz)
 = \int_0^t\ind{\xt>a}\d \xt + \half\int_0^t\delta(\xt-a)\d t.
\end{equation}
Choosing \(t=\tb\) and \(\xtz=0\) gives
\begin{equation}
u(b) - u(0) =  \int_0^{\tb}\ind{\xt>a}\d \xt + \half L_{\tb}(a).
\end{equation}
Thus
\begin{equation}
\la= 2\za + 2(u(b)-u(0))\qquad {\rm where}\qquad 
\za= \int_0^{\tb}\ind{\xt>a}\d \xt.
\end{equation}
Now define a set
\(\lbrace a_i; i=0,1,\ldots,l\rbrace\) such that 
\(a=a_0<a_1<\ldots<a_l=b\) and \(a_{i+1}-a_{i} = (b-a)/l\)
and denote the quadratic variation of \(\la\) by \(\laq\).
Then
\begin{eqnarray}
\laq &=& \lim_{l\to\infty}\sum_{i=0}^{l-1}
\left({\bf L}_{a_{i+1}}-{\bf L}_{a_{i}}\right)^2\nonumber\\
&=& \lim_{l\to\infty}\sum_{i=0}^{l-1}4
\left({\bf Z}_{a_{i+1}}-{\bf Z}_{a_{i}}\right)^2.
\label{laqv}
\end{eqnarray}
To evaluate \eqref{laqv}, we define the processes \(\st=\left(
\int_0^t \ind{a_i<\xt<a_{i+1}}\d \xt
\right)^2\).
Then \(\stb=
\left({\bf Z}_{a_{i+1}}-{\bf Z}_{a_{i}}\right)^2\).
The SDE for \(\st\) is, using \ito's formula \cite{randy,nrw},
\begin{eqnarray}
\d\st=
2\left(\int_0^t \ind{a_i<\xt<a_{i+1}}\d \xt\right)\ind{a_i<\xt<a_{i+1}}\d \xt\nonumber\\[5pt]
+ \ind{a_i<\xt<a_{i+1}}\d t.
\end{eqnarray}
Thus
\begin{eqnarray}
\stb=
2\int_0^{\tb}\left({\bf Z}_{a_{i+1}}-{\bf Z}_{a_{i}}\right)\ind{a_i<\xt<a_{i+1}}\d \xt\nonumber\\[5pt]
+ \int_0^{\tb}\ind{a_i<\xt<a_{i+1}}\d t.
\label{stb}
\end{eqnarray}
In the large-\(l\) limit, only the contributions
to \(\laq\) from the second integral in \eqref{stb} 
remain \cite{randy,nrw} and 
\begin{eqnarray}
\begin{array}{lll}
\laq &=&4\int_0^{\tb}\ind{\xt>a}\d t\\
        &=&4\int_a^{b}{\bf L}_x\d x.
\end{array}
\end{eqnarray}
Thus, defining \(z=b-a\), \({\bf L}_z\)
 satisfies the SDE \cite{nrw}
\begin{eqnarray}
\d{\bf L}_z = \left\{\begin{array}{lll}
2 {\bf L}_z^{\half} \d W_{z} - 2(f(b-z){\bf L}_z-1) \d z &z\le b,\\
2 {\bf L}_z^{\half} \d W_{z} - 2f(b-z){\bf L}_z \d z &z>b,\end{array}\right.
\end{eqnarray}
with initial condition \({\bf L}_{0}=0\)
and an absorbing boundary at \(0\).

\end{document}